\documentclass[runningheads]{llncs}
\usepackage{mathrsfs}
\usepackage[T1]{fontenc}
\usepackage{xcolor}
\usepackage{graphicx}
\usepackage{array}
\usepackage{booktabs}
\usepackage{multirow}
\usepackage{graphicx}
\usepackage{float}
\usepackage{subfig}
\usepackage{marvosym}

\begin{document}

\title{Multi-View Pre-Trained Model for Code Vulnerability Identification}

\author{Xuxiang Jiang\inst{1} \and
Yinhao Xiao\inst{2} \and
Jun Wang\inst{1} \and
Wei Zhang\inst{1}\textsuperscript{(\Letter)}}
\authorrunning{Xuxiang Jiang \textit{et al.}}

\institute{School of Computer Science and Technology, East China Normal University \\
\email{zhangwei.thu2011@gmail.com} \and
School of Information Science, Guangdong University of Finance \& Economics}
\maketitle
\begin{abstract}
Vulnerability identification is crucial for cyber security in the software-related industry.
Early identification methods require significant manual efforts in crafting features or annotating vulnerable code.
Although the recent pre-trained models alleviate this issue, they overlook the multiple rich structural information contained in the code itself.
In this paper, we propose a novel Multi-View Pre-Trained Model (MV-PTM) that encodes both sequential and multi-type structural information of the source code and uses contrastive learning to enhance code representations.
The experiments conducted on two public datasets demonstrate the superiority of MV-PTM.
In particular, MV-PTM improves GraphCodeBERT by 3.36\% on average in terms of F1 score.

\end{abstract}

\section{Introduction}

Code vulnerabilities are a major threat to the software-related industry. It is reported that the number of vulnerabilities has grown from 4,600 in 2010 to 175,477 by 2022~\footnote[1]{\url{http://cve.mitre.org/}}. The number of vulnerabilities is still rapidly increasing.


Accordingly, the field of vulnerability identification is under intensive exploration in academia.
In early-stage research, vulnerability identification methods can be categorized into three types: static analysis~\cite{Xu-ISoLA2010,Chandramohan-SIGSOFTFSE2016}, dynamic analysis~\cite{LI-ESEC/SIGSOFTFSE2017}, and machine learning methods~\cite{Grieco-CODASPY2016,Shin-IEEETrans.SoftwareEng.2011} based on hand-crafted features. Yet, a drawback of these methods restrains their performance. That is, they require vulnerability-related expertise and significant manual efforts, yielding them hard to be deployed and poorly scalable~\cite{Zhou-NeurIPS2019}. 

In later-stage research, researchers applied deep learning to address the aforementioned drawback existing in early-stage research~\cite{Lin-Proc.IEEE2020}. Some studies~\cite{Li-NDSS2018,Zhou-NeurIPS2019} leveraged several state-of-the-art deep learning techniques, e.g., LSTM and GGRN. 
A common feature of these methods is that they require large amounts of labeled data to perform supervised training and achieve better performance than conventional methods. Unfortunately, there is currently a lack of data annotated with vulnerability categories, and manually annotating data is labor-intensive. This hinders the further development of these methods in vulnerability identification.

The emergence of pre-training techniques alleviates the aforementioned problem. 
Thanks to the advancement, some pre-trained models for source code have been proposed, such as CodeBERT~\cite{Feng-EMNLP2020} and CodeT5~\cite{Wang-EMNLP2021}. However, a significant disadvantage of these methods is that they ignores rich structural information such as abstract syntax and control flow. As such, a natural research question arises: how to combine multiple structural information with pre-trained models for vulnerability identification.

To tackle this question, we propose a novel Multi-View Pre-Trained Model (MV-PTM).
Based on the pre-trained model, MV-PTM encodes both sequential information and multi-type structural information of the source code in a unified framework. Specifically, it generates representations of code under different structural information constraints. We term these representations as multiple views of source code. In this work, we use analysis tools to extract the Abstract Syntax Tree (AST), Control Flow Graph (CFG), and Data Flow Graph (DFG) of the source code and represent them as adjacency matrices that are taken as input by Structural-Aware Self-Attention Encoder to produce views containing different semantics. MV-PTM makes vulnerability predictions based on these views. In addition, MV-PTM uses contrastive learning~\cite{Oord-CoRR2018} method for representation enhancement of structural information. 

Our contributions are listed as follows:
\begin{itemize}
	\item We propose a novel approach based on the pre-trained model that learns different structural information of the source code in a unified framework, which endows our model with the capability to represent the semantics of code more accurately.
	\item We perform contrastive learning based on multiple views of code to improve the performance of code representation learning, which is demonstrated to be better at characterizing code in the experiment.
	\item MV-PTM outperforms the state of the arts significantly with an average of 3.85\% higher Accuracy and 6.80\% F-1 Score.
\end{itemize}


\begin{figure}[!t]
\includegraphics[scale=0.4]{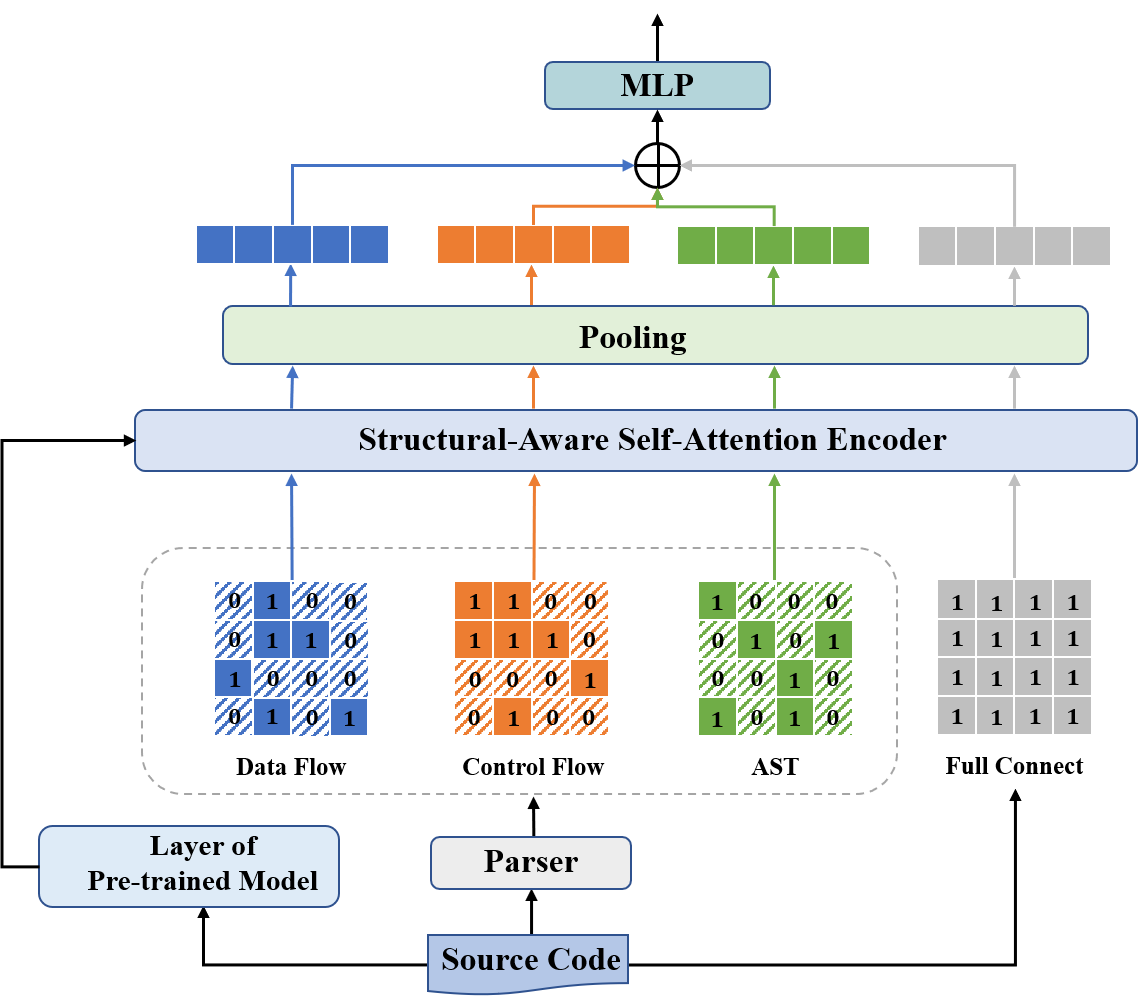}
\centering
\caption{The architecture of MV-PTM. In the adjacency matrix, 1 denotes there is an edge between corresponding nodes of code, and 0 otherwise. 
The layer of pre-trained model is used to obtain the source code embedding.
}
\label{fig1}
\vspace{-1em}
\end{figure}

\section{Methodology}

\noindent \textbf{Overview.} Fig.~\ref{fig1} shows the overview of MV-PTM. First, we use Tree-sitter~\footnote[2]{\url{https://github.com/tree-sitter/tree-sitter-c}} to parse the source codes and get the code structural graphs 
We then convert these graphs into adjacency matrices to guide the generation of multi-view code representations based on the Structural-Aware Self-Attention Encoder and pre-trained model. Afterward, the multi-view representations are fed to a Pooling Layer and Multi-layer Perceptron (MLP) for identification.


\subsection{Structural Information} 
As aforementioned, we first obtain different structural graphs: CFG, AST, and DFG. Each node in these graphs represents a program statement, and each edge represents certain structural information. 
We use an adjacency matrix $M^{n\times n}$ to represent a certain type of edges in the graph ($n$ is the number of tokens). We set $M_{i,j}=1$ if the $i$-th node and the $j$-th node are connected in the graph; Otherwise, $M_{i,j}=0$.

\begin{figure}[!t]
	\centering
	\begin{minipage}[h]{\linewidth}
	    \centering
	    \subfloat[]{\includegraphics[width=3cm,height=3.5cm]{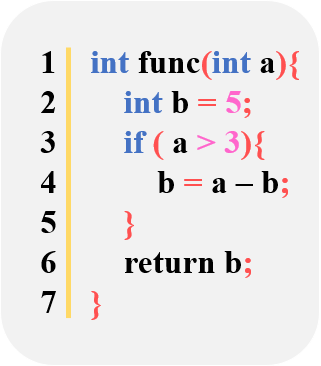}}
	    \hspace{1.5cm}
	    \subfloat[]{\includegraphics[width=5cm,height=4cm]{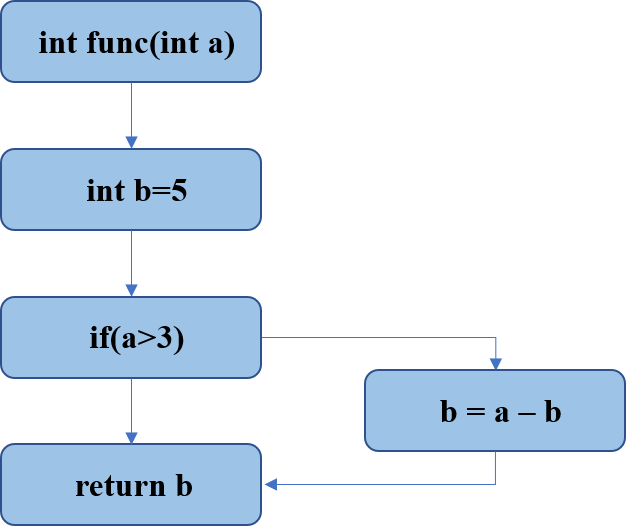}}
	    \vspace{.01in}
	    \subfloat[]{\includegraphics[width=6.5cm,height=4cm]{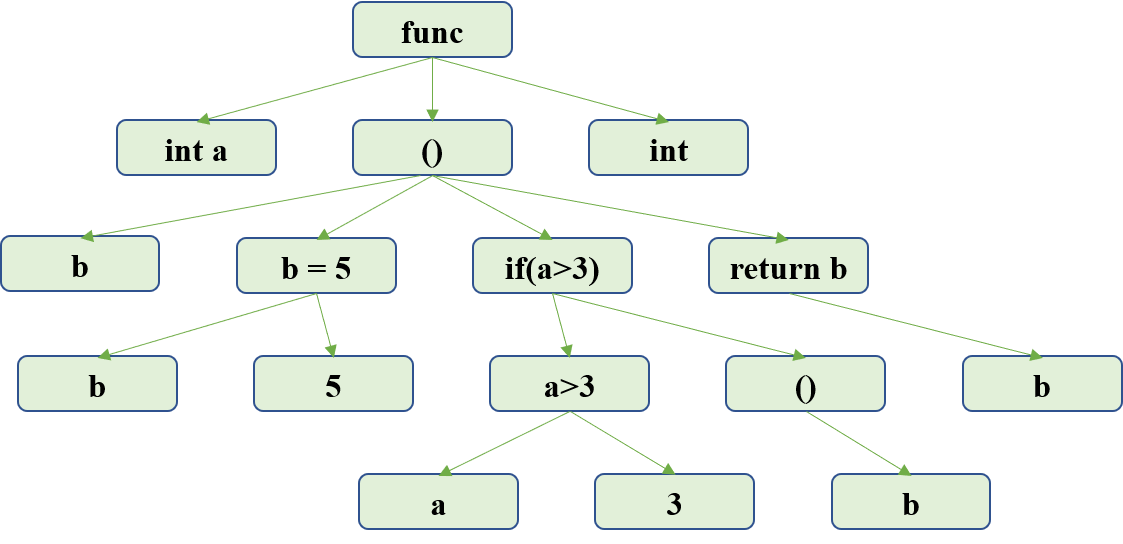}}
	    \subfloat[]{\includegraphics[width=4.5cm,height=4cm]{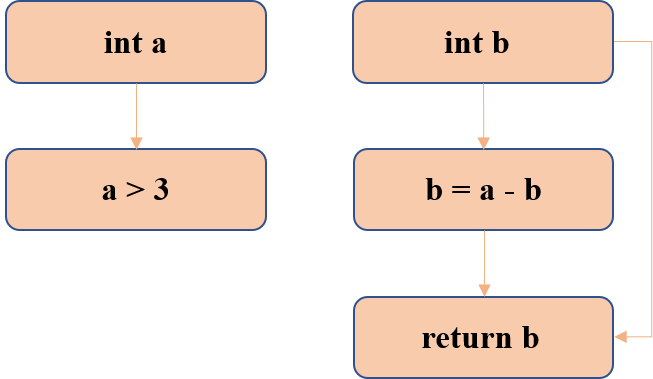}}
	\end{minipage}
	\caption{An example of different structural adjacency matrices.}
	\label{fig2}
\end{figure}

CFG is a graphical representation of the paths that are traversed during the execution of a program. For example, as shown in Fig.~\ref{fig2} (b), when the program executes the ``if (a > 3)'' statement, it decides whether ``b = a - b'' is executed according to the variable ``a''. 
AST is a tree-structured representation of the syntax structure of the source codes. Each node on the tree represents a syntactic structure. 
We use the subtrees in AST to analyze each statement in the program. Specifically, the tokens in the same statement can be connected to each other. 
DFG tracks the use of variables during program execution, including access or modification of variables. Take Fig.~\ref{fig2} (d) for instance, the variable ``a'' in ``b = a - b'' comes from ``a > 3'.

\subsection{Structure-Aware Self-Attention Encoder}
We utilize the pre-trained CodeBERT as the backbone in our approach to generate contextualized token representations, but our approach is flexible to other pre-trained models.
Taking the source code $x_i$ as an example, the representations $Z_i$ are obtained by:
\begin{equation}\label{eq0}
Z_i = \mathrm{CodeBERT}(x_i)\,.
\end{equation}

On top of the backbone, we further design Structure-Aware Self-Attention Encoder (SASA) based on the self-attention mechanism proposed by~\cite{Vaswani-NeurIPS2017}.
SASA combines the structural information matrix with the scaled dot-product attention using addition operations, given by:
\begin{equation}\label{eq1}
\mathrm{Attn}(Q,K,V,M) = \mathrm{softmax}(\frac{QK^T}{\sqrt{d_k}} + M)V\,,
\end{equation}
where $Q, K$ and $V$ denote the query, key, and value matrix, respectively, and are set by $Z_i$.
$d_k$ is the dimension of $K$ and $\mathrm{softmax}$ is Normalized Exponential Function. $M$ is the adjacency matrix generated according to the specific structure information and it can constrain what the $i$-th token can attend to when computing attention values. 
Under the constraints of adjacency matrices, SASA can generate multiple views containing different structural information.

Under our observation, we notice that there are some similar dependencies between different structural information. 
Inspired by this, we make the node representation learning on different views share the same self-attention head.
Besides, we also have the view-specific self-attention head.
To fuse the representations learned from shared heads and view-specific heads, we use linear mapping to project them into the same space.
As a whole, the SASA attention for one structural adjacency matrix (one view) is calculated as follows:
\begin{equation}\label{eq2}
\mathrm{SASA}(Q,K,V,M) = \mathrm{Cat}(H_1, H_2)W^o\,,
\end{equation}
where $H_1$ and $H_2$ correspond to the representation learned from the shared self-attention head and view-specific self-attention head, as shown in Eq.~\ref{eq1}.
$\mathrm{Cat}$ means the concatenation operation.

\subsection{Multi-view Contrastive Learning}
To enhance the representations learned from different structural information, we regard each type of information as one view and perform Contrastive Learning.
This is motivated by the fact that different views of the same piece of code have some correlations and tend to cluster together in the semantic space. 

To realize contrastive learning, we consider different views of the same code as positive pairs and those of different codes as negative pairs. 
The loss function w.r.t. contrastive learning is expressed as:
\begin{equation}\label{eq3}
\mathcal{L}_{CONTRA}= \psi_{ast} + \psi_{dfg} + \psi_{cfg}\,,
\end{equation}
where $\psi_{ast}$ takes AST as the Matched Structural view, and it is analogous to $\psi_{dfg}$ and $\psi_{cfg}$. $\psi$ is Normalized Temperature Scaled Cross Entropy Loss~\cite{Chen-ICML2020}.

\begin{figure}[h]
\includegraphics[scale=0.5]{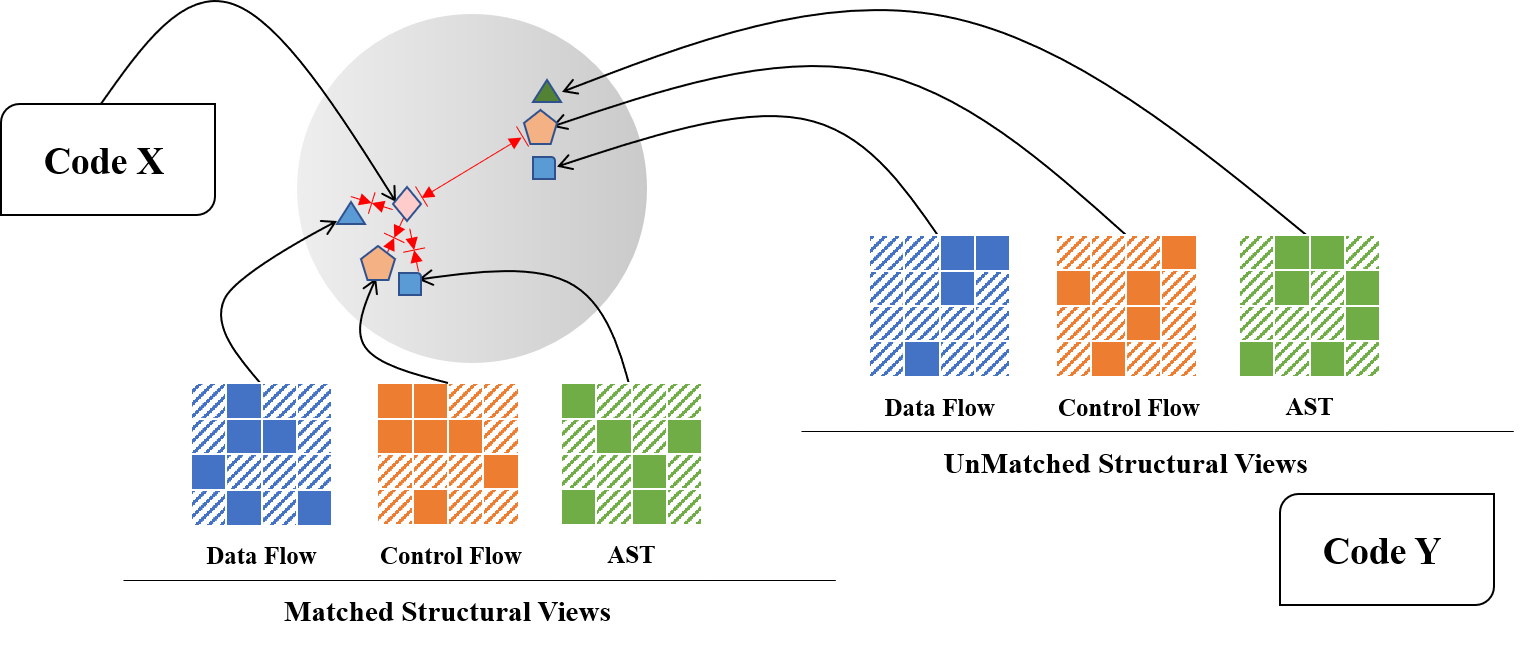}
\centering
\caption{Diamonds, triangles, pentagons, and squares correspond to code sequence, DFG, CFG, and AST, respectively. The circle denotes the semantic space.}
\label{fig3}
\end{figure}

\subsection{Training Loss}
We leverage Cross Entropy for training the main task, i.e., vulnerability identification, 
and the total loss for fine-tuning MV-PTM is given by:
\begin{equation}\label{eq6}
\mathcal{L}=\mathcal{L}_{CLS}+\lambda\mathcal{L}_{CONTRA}\,,
\end{equation}
where $\lambda$ is the hyper parameter and we set $\lambda = 1$ in the experiments.
\section{Experiments}
\subsection{Experimental Setup}

\noindent \textbf{Datasets.}
We evaluate our approach on two C-language datasets used in previous studies~\cite{Zhou-NeurIPS2019,Wang-EMNLP2021}, which contain manually-labeled functions collected from open-source projects FFmpeg and QEMU. 

Since some code snippets in the dataset exceed the length limit of CodeBERT, we discarded the codes that have more than 512 tokens. For long code segments, the recognition accuracy of MV-PTM is not ideal.

\begin{table}[!h]
\centering
\caption{Statistics of the datasets.}\label{tab1}
\begin{tabular}{c c c}
\toprule[1.5pt]
\makebox[0.1\textwidth][c]{} &  \makebox[0.2\textwidth][c]{FFmpeg} & \makebox[0.2\textwidth][c]{QEMU}\\
\hline
Training Set &  3958 & 10903\\
Validation Set &  462 & 1378\\
Test Set & 499 & 1319\\
Total & 4919 & 13600\\
\hline
Average Length & 274.5 & 325.3\\
\bottomrule[1.5pt]
\end{tabular}
\end{table}

\noindent \textbf{Baselines.}
We choose the following six methods as the baselines since they represent the most up-to-date vulnerability identification mechanisms:

\textbf{VulDeePecker}~\cite{Li-NDSS2018}: It turns the source codes into a token sequence. The initial embeddings of tokens are trained via Word2Vec~\cite{Mikolov-NeurIPS2013}.

\textbf{CNN}~\cite{Russell-ICMLA2018}: It models the source codes as natural language and applies CNN to extract features from the code. The embedding initialization is the same as that of VulDeePecker.

\textbf{Devign}~\cite{Zhou-NeurIPS2019}: It represents the source code with the code property graph (CPG) which integrates all syntax and dependency semantics.
Based on the graph, it uses Gated Graph Recurrent Network\cite{Li-ICLR2015} for graph-level classification.

\textbf{SELFATT}~\cite{Vaswani-NeurIPS2017}: Similar to~\cite{Li-NDSS2018}, it takes the source code as sequences and exploits the multi-head attention mechanism for code representation learning.

\textbf{CodeBERT}~\cite{Feng-EMNLP2020}: It is a pre-trained model for programming language which has achieved acceptable performance on many code-related tasks such as code search and code documentation generation.

\textbf{GraphCodeBERT}~\cite{Guo-ICLR2021}: It is the first pre-trained model that leverages code structure to learn code representation to improve code understanding.


\subsection{Experimental Results}
\noindent \textbf{Performance comparison.}
As shown in Table~\ref{tab2}, MV-PTM outperforms all baseline methods on both two datasets. According to the experimental results, we summarize the following findings:

\textbf{The local and structural characteristics of the code can improve the performance of vulnerability identification.} Comparing CNN with VulDeePecker, we can find that the Accuracy is significantly improved in both two datasets, implying that the local characteristics learned by CNN are indeed helpful for vulnerability identification. GraphCodeBERT outperforms CodeBERT with an average of 1.64\% higher Accuracy and 3.44\% F-1 Score.


\textbf{MV-PTM performs best among all methods.} MV-PTM has further improved its performance based on CodeBERT. It is noteworthy that the F-1 Score of baselines on the QEMU dataset is not ideal, while MV-PTM raised the F-1 Score to 0.7049. 

\begin{table}[!t]
\centering
\setlength{\tabcolsep}{3.5mm}{
	\caption{Experimental results on the two datasets.}\label{tab2}
	\begin{tabular}{*{5}{c}} 
		\toprule[1.5pt]
		\multirow{2}{*}{\textbf{Methods}} & \multicolumn{2}{c}{\textbf{FFmpeg}} & \multicolumn{2}{c}{\textbf{QEMU}} \\ 
		\cmidrule(lr){2-3}\cmidrule(lr){4-5}
		& Accuracy & F-1 Score & Accuracy &  F-1 Score  \\
		\midrule[0.75pt]
		{VulDeePecker}\cite{Li-NDSS2018} & 0.5622 & 0.5923 & 0.5956 & 0.5644\\
		{CNN}\cite{Russell-ICMLA2018} & 0.6032 & 0.6278 & 0.6482 & 0.3974\\
		{Devign}\cite{Zhou-NeurIPS2019} & 0.5904 & 0.6015 & 0.6039 & 0.3244\\
		{SELFATT}\cite{Vaswani-NeurIPS2017} & 0.6152 & 0.6323 & 0.6361 & 0.3701\\
		{CodeBERT}\cite{Feng-EMNLP2020} & 0.6353 & 0.6431 & 0.6907 & 0.6102\\
		{GraphCodeBERT}\cite{Guo-ICLR2021} & 0.6613 & 0.6724 & 0.6975 & 0.6497\\
		MV-PTM  & \textbf{0.6874} & \textbf{0.6843} & \textbf{0.7149} & \textbf{0.7049}\\
		\bottomrule[1.5pt]
	\end{tabular} 
}
\end{table}

\subsection{Ablation study}
In this section, we verify the effectiveness of the three structural information and contrastive learning methods used in our work according to the results of the ablation study. Table~\ref{tab3} shows the experimental results.

\begin{table}[!h]
\vspace{-1.5em}
\centering
\setlength{\tabcolsep}{3.5mm}{
	\caption{Effect of structural information and contrastive learning.}\label{tab3}
	\begin{tabular}{*{5}{c}} 
		\toprule[1.5pt]
		\multirow{2}{*}{\textbf{Methods}} & \multicolumn{2}{c}{\textbf{FFmpeg}} & \multicolumn{2}{c}{\textbf{QEMU}} \\ 
		\cmidrule(lr){2-3}\cmidrule(lr){4-5}
		& Accuracy & F-1 Score & Accuracy &  F-1 Score  \\
		\midrule[0.75pt]
		MV-PTM  & \textbf{0.6874} & \textbf{0.6843} & \textbf{0.7149} & \textbf{0.7049}\\
		MV-PTM w/o CFG & 0.6553 & 0.6643 &  0.6983 & 0.6865\\		
		MV-PTM w/o AST & 0.6573 & 0.6674 &  0.7036 & 0.6845\\
		MV-PTM w/o DFG & 0.6513 & 0.6683 &  0.6990 & 0.6892\\
		MV-PTM w/o Contrastive & 0.6693 & 0.6845 &  0.7005 & 0.6688\\
		\bottomrule[1.5pt]
	\end{tabular} 
}
\end{table}

\textbf{Structural information improves the performance of the model.} It can be observed from Table~\ref{tab3} that when any kind of structural information is removed, the performances of the model on both datasets decrease significantly. 

\textbf{Contrastive Learning can learn a more accurate multi-view representation of source code.} We find that after removing the contrastive learning module, the performance of the model also decreases to a certain extent. This phenomenon implies that the contrastive learning method we proposed can make the model learn different code structure information more effectively.

\section{Related Work}
\noindent\textbf{Vulnerability Identification.}
In academia, there usually are rule-based and learning-based methods for vulnerability identification. Rule-Based methods are widely explored in academia. SUTURE~\cite{Zhang-CCS2021} is a static analysis method, which is capable of identifying high-order vulnerabilities in OS kernels. 
Learning-Based methods are a novel research direction that attracts much attention. VulDeePecker~\cite{Li-NDSS2018} is an LSTM-based model, which represents the source code as vectors. 


\noindent\textbf{Pre-trained Model for Programming Languages.}
CodeBERT proposed in~\cite{Feng-EMNLP2020} is a bimodal model for programming language and natural language trained by Masked Language Modeling and Replaced Token Detection. GraphCodeBERT ~\cite{Guo-ICLR2021} considers the inherent structure of code by Edge Prediction and Node Alignment to support tasks like code clone detection~\cite{Sheneamer-IEEEAccess2021,Huang-CoRR2021,Ji-Int.J.Softw.Eng.Knowl.Eng.2021}. 
Compared to CodeBERT, MV-PTM improves the accuracy by an average of 3.81\% at the cost of 15\% additional parameters and 25\% training time, which is within acceptable limits.

\noindent\textbf{Contrastive Learning.}
Contrastive learning is usually conducted in an unsupervised manner by increasing the similarity between the representations of positive pairs and decreasing the similarity between the representations of negative pairs~\cite{He0WXG-CVPR20,TangDZ22}.
Data augmentation is a commonly-used technique to construct positive pairs, including rotation, scaling, and cropping in computer vision~\cite{Chen-ICML2020} and dropout in NLP~\cite{Gao-EMNLP2021}.


\section{Conclusion}
In this paper, we propose MV-PTM, a pre-trained based model which uses structural information including AST, DFG, and CFG, to obtain multiple views of the source code. Besides, we introduce an auxiliary task based on contrastive learning to improve the performance of code representation. The experiments on two datasets demonstrate that structural information and contrastive learning are effective for vulnerability identification. In the future, we plan to introduce the knowledge graph to generate reasonable explanations for the identified vulnerabilities.

\section*{Acknowledgment}
This study was supported in part by the National Key R\&D Program of China (2019YFB2102600), the National Natural Science Foundation of China (62002067), and the Guangzhou Youth Talent Program of Science (QT20220101174).

\bibliographystyle{splncs04}
\bibliography{reference}
\end{document}